\documentclass[prl,twocolumn,showpacs,amsmath,amssymb,superscriptaddress]{revtex4}

\usepackage{graphicx,amsmath}

\newcommand{\unit}[1]{\,{\rm #1}}
\newcommand{\sub}[1]{_{\rm #1}}
\newcommand{\subup}[1]{^{\rm #1}}
\newcommand{\e}[1]{{\rm e}^{#1}}
\newcommand{\vectorC}{\boldsymbol {\mathcal{C}}}
\newcommand{\vctr}[1]{{\bf {#1}}}
\newcommand{\hc}{{\rm h.\,c.}}
\newcommand{\ann}[1]{#1^{\phantom{\dagger}}}
\newcommand{\cre}[1]{#1^{\dagger}}
\newcommand{\citeq}[1]{Eq.\,(\ref{#1})}
\newcommand{\resp}{{\textit{resp. }}}
\begin{document}

\title{Spin-Electric Coupling in Molecular Magnets}

\author{Mircea Trif}
\affiliation{Department of Physics, University of Basel,
             Klingelbergstrasse 82, CH-4056 Basel, Switzerland}
\author{Filippo Troiani}
\affiliation{CNR-INFM National Research Center S3  c/o Dipartimento
di Fisica via G. Campi 213/A, 41100, Modena, Italy}
\author{Dimitrije Stepanenko}
\affiliation{Department of Physics, University of Basel,
             Klingelbergstrasse 82, CH-4056 Basel, Switzerland}
\author{Daniel Loss}
\affiliation{Department of Physics, University of Basel,
             Klingelbergstrasse 82, CH-4056 Basel, Switzerland}

\date{\today}

\begin{abstract}
   
We study the triangular antiferromagnet Cu$_3$ in external electric
fields, using symmetry group arguments and a Hubbard model approach.
We identify a spin-electric coupling caused by an interplay between
spin exchange,  spin-orbit interaction, and the chirality of the
underlying spin texture of the molecular magnet. This coupling allows for the
electric control of the spin (qubit) states, e.g. by using an STM tip or a
microwave cavity. We propose an experimental test for identifying
molecular magnets exhibiting spin-electric effects.

\end{abstract}


\maketitle

Single-molecule magnets (SMMs) \cite{gatteschi} have emerged as a
fertile testing ground for investigating quantum effects at the
nanoscale, such as tunneling of magnetization
\cite{FST+96,TLB+96,BKW00}, or
coherent charge transport \cite{RWH+06,LM06,LL07}, or the decoherence and the transition from quantum to classical
behavior \cite{ARM+07}.  SMMs with antiferromagnetic coupling
between neighboring spins are especially promising for the encoding
and manipulation of quantum information
\cite{LL01,meier03,troiani05b,lehma07}, 
for they act as effective
two-level systems, while providing additional auxiliary states that can
be exploited for performing quantum gates, even in the presence of
untunable couplings between the qubits \cite{troiani05a}.  Intra- and
inter-molecular couplings of SMMs can be engineered by molecular and
supra-molecular chemistry \cite{affro05}, enabling 
a bottom-up design of molecule-based devices \cite{bogan08}.

While the properties of SMMs can be modfied during
the synthesis, control on the time scales required
for quantum information processing remains a challenge.  The standard
spin-control technique is electron spin resonance (ESR) driven by ac
magnetic fields ${ B}_{ac}(t)$ \cite{ARM+07,BGT+07,BFC+08,BG89}.  For manipulation on the time
scale of $1\unit{ns}$ (Rabi frequency $\Omega_R \sim 10^9\unit{s}^{-1}$)
 $  {B}_{ac}$ should be of the order of
$10^{-2}\unit{T}$, which, however, is difficult to achieve.  The spatial resolution of $1\unit{nm}$,
required for addressing a single molecule, is also prohibitively
small.  At these spatial and temporal scales, the electric control is
preferable, because strong electric fields can be applied to small
regions by using, for example,  STM tips
\cite{HFS03,HLH06,BFB+08arXiv}. Also, the quantized electric field
inside a microwave cavity can be used \cite{WSB+04,BI06,ADD+06,TGL08}
to control single qubits and to induce  coupling between them even if they are far apart.

Here we identify and study an efficient spin-electric coupling
mechanism in SMMs which is based on an interplay of spin exchange,
spin-orbit interaction (SOI), and lack of inversion symmetry.
Spin-electric effects induced solely by SOI \cite{rashb03}  have been
proposed \cite{GBL06} and experimentally demonstrated \cite{NKN+07} in
quantum dots. However,  these SOI effects scale with the system 
size $L$ as $L^3$ \cite{GBL06}, making
them  irrelevant  for the much smaller SMMs. Thus, additional ingredients--such as broken symmetries--
must be present in SMMs for an efficient coupling  between spin and applied electric field.

In the following, we demonstrate the possibility of such spin-electric effects in SMMs by
focusing on a specific example, namely an equilateral spin triangle,
Cu$_3$ \cite{choi06}.  In this SMM, the low energy states exhibit
a chiral spin texture and, due to the absence of inversion symmetry,
electric fields 
couple states of opposite
chirality.  Moreover, SOI couples the chirality to the total
spin, and thus  an effective spin-electric interaction eventually emerges.

{\textit{Spin-electric coupling.}}  The low-energy
states of Cu$_3$  can be described in terms of an effective spin
Hamiltonian.  There, three spins-1/2 ${\bf s}_i$ (one for each Cu$^{2+}$
ion) are coupled through Heisenberg and Dzyaloshinski-Moriya
interaction, accounting  for (super-) exchange and spin orbit
interaction, \resp  \cite{choi06}, 
\begin{equation}
\label{DM}
H_0 = 
\sum_{i=1}^3 J_{i\,i+1} {\bf s}_i \cdot {\bf s}_{i+1}
+ 
\sum_{i=1}^{3} {\bf D}_{i\,i+1} \cdot {\bf s}_i \times {\bf s}_{i+1}.
\end{equation}
The  D$_{3h}$ symmetry of the triangle implies several relations between  the
coupling constants \cite{tsukerblat}.  Since $J_{pq}\sim 5\unit{K}$
and $|\vctr{D}_{pq}|\sim 0.5\unit{K}$, the Heisenberg  terms determine
the gross structure of the energy spectrum,
and the Dzyaloshinski-Moriya terms the fine one.  In
particular,  due to the antiferromagnetic  coupling  ($
J_{pq} > 0 $), the ground state multiplet corresponds to total spin
$S=1/2$, and the gap to the first excited $S=3/2$ quadruplet is $ \Delta \sub{H}
\equiv 3 J / 2 $.  The $S=1/2$ subspace is spanned by the
symmetry-adapted  states $ | \chi , M \rangle $, {\it i.e.},
\begin{eqnarray}
\label{eq:pmplus}
| \pm 1,+1/2\rangle  & \equiv &  ( | \!\downarrow\uparrow\uparrow
\rangle + \epsilon_{\pm}  | \!\uparrow\downarrow\uparrow \rangle
+ \epsilon_{\mp}\, | \!\uparrow\uparrow\downarrow \rangle )/
\sqrt{3} ,
\\
\label{eq:pmminus}
| \pm 1,-1/2\rangle  & \equiv & ( | \!\uparrow\downarrow\downarrow
\rangle + \epsilon_{\pm}  | \!\downarrow\uparrow\downarrow \rangle
+ \epsilon_{\mp}\, | \!\downarrow\downarrow\uparrow \rangle )/
\sqrt{3},
\end{eqnarray}
with $\epsilon_{\pm}= \e{\pm i2\pi/3}$, that are simultaneous
eigenstates of the chirality operator ${\mathcal{C}}_z$ and of $S_z$ (total spin),
for the respective eigenvalues $\chi$ and $M$.  
Here, we have introduced the
chirality  vector $\vectorC $ 
with components
\begin{align}
\label{eq:cx}
\mathcal{C}_x &= (-2/3) ( \vctr{s}_1 \cdot \vctr{s}_2 - 2 \vctr{s}_2 \cdot \vctr{s} _3 + \vctr{s}_3 \cdot \vctr{s}_1 ),
\\
\label{eq:cy}
\mathcal{C}_y &= (2/\sqrt{3}) ( \vctr{s}_1 \cdot \vctr{s}_2 - \vctr{s}_3 \cdot \vctr{s}_1 ),
\\
\label{eq:cz}
\mathcal{C}_z &= (4/\sqrt{3}) \vctr{s}_1 \cdot ( \vctr{s}_2 \times \vctr{s}_3 ).
\end{align}
They satisfy $[{\mathcal C}_k,{\mathcal C}_l]=i 2\epsilon_{kln} {\mathcal C}_n$ and
  $[{\mathcal C}_k, S_l]=0$, 
and  act as Pauli matrices in the $|\chi = \pm 1\rangle$ bases.  The states $ | \chi , M
\rangle $ are split by the SOI (see below) into two Kramers
doublets,  $| \pm 1, \pm 1/2\rangle $ and  $| \pm 1, \mp 1/2\rangle $. 
\begin{figure}[t]
\begin{center}
\includegraphics[width=8.5cm]{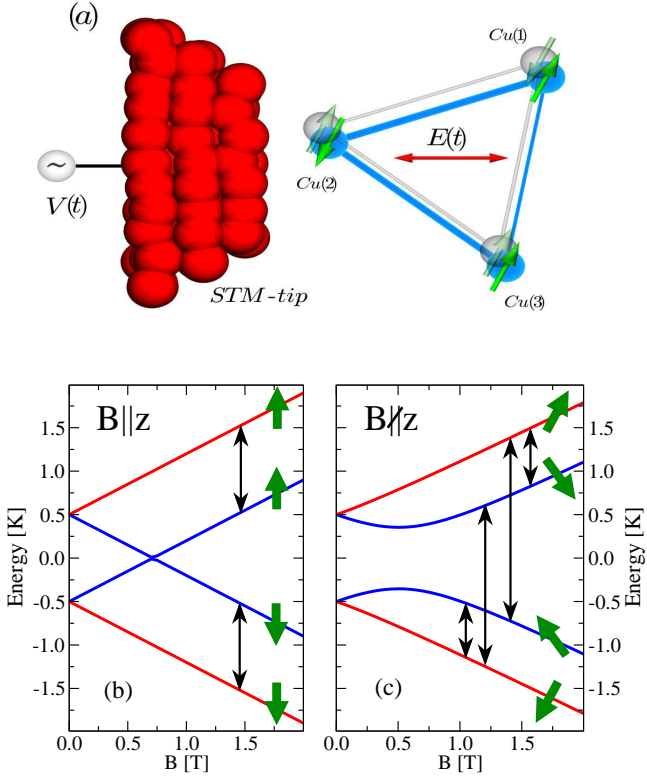}\vspace{0.1cm}
\includegraphics[width=8.5cm]{transitions_1.eps}
\caption{\label{fig:levels_triangle}(Color online) (a) 
Cu$_3$-triangle exposed to an electric field ${\bf E}$  created by, e.g., an STM-tip.
 For ${\bf E}=0$, the
  exchange couplings in the molecule are equal (light triangle).  A finite ${\bf E}$ 
  affects the (super-) exchange
  coupling in a directional way (dark triangle).  
(b, c) Low-energy $S=1/2$ states of  Cu$_3$ in a magnetic field ${\bf B}$, with the zero-field
  SOI splitting $\Delta \sub{SO}\!\!\! =\!\! 1\unit{K}$.  Light (red) and dark (blue) lines correspond to states with $\chi =\!
  +1(-1)$.  If $\vctr{B}\parallel z$  (b), the 
transitions induced by ${\bf E}$  (thin arrows) conserve $S_z$; 
for ${\bf B}\nparallel z$ (c), 
these transitions result in a change of spin orientation (thick arrows). }
\end{center}
\end{figure}

Next, we study the effect of an electric field ${\bf E}$ on the Cu$_3$-spins using
general symmetry group arguments.
The low-energy
molecular states form ${^2\! E'}$ and ${^4\!\! A_2'}$ irreducible
representations (IRs) of D$_{3h}$, with the $S=1/2$-states lowest in
energy and transforming as two $E'$ IRs, while the higher energy
$S=3/2$-states span the four $A_2'$ IRs \cite{tsukerblat,choi06}.
The chiral states, Eqs.\,(\ref{eq:pmplus},\ref{eq:pmminus}), form the standard bases of the 
$E'$-representations, with
$|E'_\pm, S_z\rangle = |\chi =\pm 1, S_z \rangle$.  An electric field
$\vctr{E}$ couples to Cu$_3$ via  $e\vctr{E}\cdot
\vctr{R}$, where  $e$ is the electron charge, and
$\vctr{R} = \sum_{j=1}^3 \vctr{r}_j$  transforms
according to the reducible vector representation $D\sub{v} = A_2'
\oplus E'$.  The $A_2'$ IR acts on $Z$ and the two-dimensional IR
$E'$ acts on the (X,Y)-components in the Cu$_3$-plane.  The only non-zero matrix
elements of $\vctr{R}$ in our basis are $e\langle
E_+', S_z | X_- | E_-', S_z \rangle = e\langle E_-',S_z |
X_+ | E_+',S_z \rangle = 2id $, with $d$ real denoting the electric dipole
coupling, and $X_{\pm}=\pm X+iY$ being the irreducible
tensor components of $\vctr{R}$.  The zero matrix elements imply that the $S=3/2$-states are unaffected  in
first order in $\vctr{E}$, and $E_z$ has no effect on the spins at all. The resulting coupling between the $\vctr{E}$-field and  chirality  is described by  $\delta
H\sub{E}=d \vctr{E}' \cdot \vectorC _{\parallel}$, where
$\vctr{E}'={\mathcal R}_z(\phi) \vctr{E}$ is  rotated by
$\phi=7\pi/6 -2\theta$ about  $z$, and  $\vectorC _{\parallel} = ({\mathcal C}_x,
{\mathcal C}_y,0)$.

To emphasize that the spin-electric effect derived above is based on
exchange, we reinterpret our results in terms of spin interactions.  In
an equilateral triangle, and in the absence of electric field, the
spin Hamiltonian is given by \citeq{DM}
with equal exchange couplings $J_{i,i+1}\equiv J$. Using  then
Eqs.\,(\ref{eq:cx}) and (\ref{eq:cy}), we find 
\begin{equation}
\delta H\sub{E} =
\frac{4dE}{3} \sum_{i=1}^{3} \sin [ 2(1-i)\pi/3+\theta ] \, {\bf s}_i
\cdot {\bf s}_{i+1},
\label{eq:hespin}
\end{equation}
where $\theta$ is the angle between an in-plane ${\bf E}$-field  and
the vector ${\bf r}_{12}$ pointing from site $1$ to  $2$.
This form of $\delta H\sub{E}$ shows that the  $\vctr{E}$-field lowers the
symmetry by introducing direction-dependent corrections to the
exchange couplings $J_{i\,i+1}$.  E.g., if $ \theta =
\pi/2 $, $\delta J_{23} = \delta J_{31} \neq \delta J_{12}$.  We mention that the
eigenstates can be labeled by a partial spin sum quantum number
$S_{12}=0,1$ of $ ( \vctr{s}_1 + \vctr{s}_2 )^2$.

While the specific form of  $d$ in
Eq.\,(\ref{eq:hespin}) depends on  microscopic details,
the structure of $\delta H\sub{E}$
relies
solely on symmetry arguments 
and the lack of inversion symmetry is crucial for  the spin-electric coupling.  Indeed,  in inversion-symmetric
SMMs, the  electric field, which is odd under inversion, and the spin, which is even, can only couple through transitions between
orbitals of opposite parity.  When the orbital state is quenched, as
in  the spin Hamiltonian description, there are no orbitals
of opposite parity. Thus, in the symmetric case, there is no spin-electric effect up to linear order in  $\vctr{E}$.
 
Next, we turn to the SOI. The most general form of SOI allowed by the $D_{3h}$
symmetry reads,
$H\sub{SO} = 
\sum_{i=1}^{3} 
[ \lambda^{\parallel}\sub{SO} T_{A_2''} s_z^i +
  \lambda^{\perp}\sub{SO} ( T_{E_+''}s_-^i + T_{E_-''} s_+^i ) ]$, 
where $\lambda^{\perp}\sub{SO}(\lambda^{\parallel}\sub{SO})$ is the
effective SOI coupling constant for the
$A_2''$- ($E_{\pm}''$-) irreducible representation, and 
$T_{A_2''}$ ($T_{E_\pm''}$) is the corresponding irreducible tensor
operator \cite{tsukerblat}.  Using again
symmetry group arguments,
we find that the SOI Hamiltonian acting in the $S=1/2$ subspace reads  $\delta H \sub {SO} = \Delta \sub{SO} {\mathcal C}_z S_z$, where
 $\Delta \sub{SO} = \lambda \sub{SO}^{\parallel}$.  Using Eq. ~(\ref{eq:cz}), this general form can be
reduced  to the Dzyaloshinski-Moriya interaction given in \citeq{DM}. 

The coupling  to a  magnetic field $\vctr{B}$ is given by  
$\vctr{B} \cdot {\bar{\bar{g}}} \cdot \vctr{S}$, with the Bohr magneton absorbed in the 
 ${\bar{\bar{g}}}$-tensor.  Due to the $D_{3h}$-symmetry, ${\bar{\bar{g}}}$ is diagonal
with components
$g_{xx}=g_{yy}=g_{\perp}$ in the Cu$_3$-plane and 
$g_{zz}=g_{\parallel}$ normal to it.

Combining  $\delta H\sub{E}$ and $\delta
H\sub{SO}$, we finally obtain  the effective
low-energy Hamiltonian in the presence of SOI and  electric and magnetic fields,
\begin{equation}
H\subup{spin}\sub{eff} = 
\Delta \sub{SO} \mathcal{C}_z\,S_z 
+ 
g_{\perp}\, \vctr{B}_{\perp} \cdot \vctr{S} +
g_{\parallel} B_z S_z +
d \, \vctr{E}' \cdot {\vectorC}_{\parallel} .
\label{central}
\end{equation}
From this we see that  an in-plane $\vctr{E}$-field  causes rotations of the chirality pseudospin.  To illustrate the role of ${\bf B}$, we focus on the case
${\bf E} \parallel {\bf r}_{31}$,  corresponding to $\delta H\sub{E} = - d
E\mathcal{C}_x\, $.  For $ {\bf B} \parallel z$, the
eigenstates coincide with those of $S_z$,  and thus  $\vctr{E}$ will not induce transitions between $|\pm, 1/2\rangle$ and $|\pm,-1/2\rangle$, but will do so in subspaces of  given $M$, see  Fig.\,\ref{fig:levels_triangle}. 
For $\vctr{B}\nparallel z$, instead, the system eigenstates for ${\bf E}=0$
are no longer  eigenstates of $S_z$, and thus the electric field induced
transitions result in spin flips, see Fig.\,\ref{fig:levels_triangle}(c).

\begin{figure}[t]
 \begin{center}
 \includegraphics[width=8.5cm,angle=0]{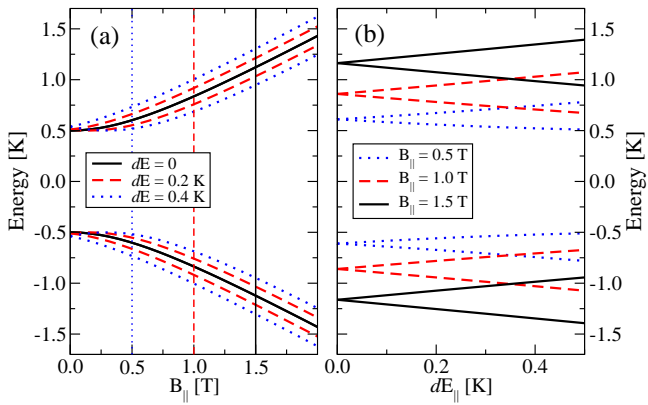}
   \caption{\label{fig:sbs}(Color online) Low-energy spectrum of the Cu$_3$
   molecule.  (a) Energy levels of the Cu$_3$ molecule in an
   in-plane magnetic field $B_{\parallel}$ (black solid
   line), split as a static in-plane electric field
   $E_{\parallel}$ is turned
   on (dashed red line and dotted blue line). (b) The electric dipole coupling
   $d$ can be determined from the slope of ESR lines
   in a constant magnetic field [vertical lines in (a)]
   as a function of $E_{\parallel}$.} 
\end{center} 
\end{figure}

We emphasize that the electric dipole coupling $d$ in
Eq.\,(\ref{central}) can be directly accessed in experiments, e.g., by standard ESR measurements 
\cite{BG89} in static electric fields, see
Fig.~\ref{fig:sbs}.  While a microscopic evaluation of $d$ requires an
ab-initio approach, which is beyond the scope of this work, we can
estimate $d$, $|\vctr{E}|$ and the spin-manipulation (Rabi) time
resulting from \citeq{central} as follows.
For $d$ between $d\sub{min}=10^{-4}e
R_{12}$ and $d\sub{max}= e R_{12}$ 
and for  $E\approx 10^2\unit{kV/cm}$,
obtainable near an STM tip, the Rabi time is $\tau \sub{Rabi}\approx
0.1-10^{3}\unit{ps}$.  The
condition $dE \ll \Delta \sub{H}$ for the validity of $H_{\rm eff}^{\rm spin}$ in \citeq{central} provides another lower bound on the spin-manipulation time, namely
$\tau \sub{Rabi} \subup{min}\approx 10\unit{ps}$.

{\textit{Hubbard approach.}}  In order to gain further insight into the interplay between the exchange
interaction and the electric field, we introduce an $N_s$-site
Hubbard model of the  triangular spin chain. The corresponding
Hamiltonian reads,
$H\sub{H} =  \sum_{i,\sigma} \left[({U_i}/{2}) n_{i,\sigma}
  n_{i,-\sigma} + \epsilon_i \, n_{i,\sigma}  + (t_{i\,i+1}
  \cre{c}_{i,\sigma} \ann{c}_{i+1,\sigma} +\hc)\right]$,
where $U_i$ is the repulsion on site $i$, $t_{i\,i+1}$ the hopping matrix 
element, $\sigma=\,\uparrow,\downarrow$, and $\sum_{i,\sigma}  n^\sigma_i =
N\sub{e} $. 
The coupling of the system to the electric field $ {\bf E} $ is 
\begin{equation}
\label{eq:helectric}
H\sub{E} =  e {\bf E} \cdot
\sum_{i,\sigma} 
\left[ n_{i,\sigma} {\bf r}_i + (\widetilde{{\bf r}}_{i\,i+1}
\cre{c}_{i,\sigma} \ann{c}_{i+1,\sigma} + \hc)\right].
\end{equation}
In the single-site terms, the expectation value of the electron position ${\bf r}$ in 
the Wannier state $|\phi_i\rangle$ is identified with the corresponding 
ion position, $ \langle \phi_i | {\bf r} | \phi_i \rangle \simeq 
{\bf r}_i $. The two-site terms describe the electric-field assisted 
hopping of electrons between neighboring sites, with 
$ {\bf \widetilde{r}}_{ii+1} = 
\langle \phi_i | {\bf r} | \phi_{i+1} \rangle = 
\alpha^{\parallel}_{i\,i+1} \vctr{r}_{i\,i+1} + 
\alpha^{\perp} \vctr{e}_z \times \vctr{r}_{i\,i+1}$.
We now focus on the two 
main mechanisms giving antiferromagnetic coupling, 
namely direct exchange and superexchange (models A and
B, Fig. 2(a)).  In both cases, the low-energy subspace ($\mathcal{S}_0$) is defined by the 
states ($ |\alpha^0 \rangle $) where the magnetic ions at the triangle 
vertices are singly occupied.
For $E=0$, the projection of these states onto $\mathcal{S}_0$ 
($| \Psi_{1-8}^0 \rangle $) coincides with the $S=1/2$ and $S=3/2$ 
eigenstates of the Heisenberg Hamiltonian. The degeneracy in the $S=1/2$
multiplet is lifted by an electric field.

%
\begin{figure} 
 \begin{center}
 \includegraphics[width=8.5cm,angle=0]{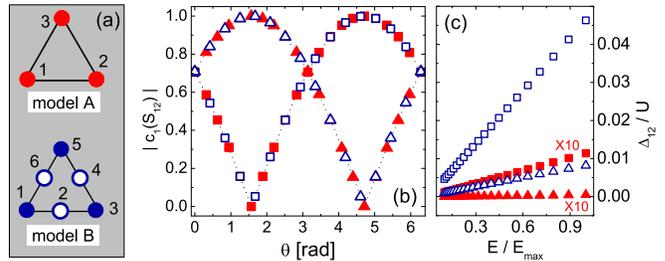}
   \caption{\label{fig2} (Color online) (a) Hubbard models A and B 
of the spin triangle. 
Model A: $ N_e \!\! = \!\! N_s \!\! = \!\! 3 $, $ t_i \!\! = \!\! t $,
$ U_i \!\! = \!\! U $, and $\epsilon_i \!\! = \!\! 
\alpha^\parallel_{i\,i+1} \!\! = \!\! 0$. 
Model B: $N_e \!\! = \!\! 9$, $N_s \!\! = \!\! 6$, 
$\alpha^\parallel_{i\,i+1} \!\! = \!\! 0$, 
$\alpha^\perp_{i\,i+1} \!\! = \!\! \alpha $, 
$ t_i\!\! = \!\! t$, 
$ \epsilon_{3k-2}\! -\! \epsilon_{3k'-1}\! =\! \epsilon $, 
$ U_{3k-2}\! -\! U_{3k'-1}\! =\! U $ ($ k , k'\! \!  =\! \!  1 , 2 , 3 $).
(b) Overlap between the projected ground state of $H_H+H_E$
($ | \Psi^0_1 \rangle $), and the eigenstates of 
$ {\bf s}_{1} \cdot {\bf s}_{2} $ corresponding to $S_{12}\!\! =\!\! 0$ 
(squares) and $ S_{12}\!\!=\!\!1 $ (triangles), as function of
the angle $\theta$ between the triangle side 1-2 and an in-plane $\vctr{E}$. 
The filled (empty)
symbols correspond to the A (B) model, whereas the dotted 
lines give the components of the $\delta H\sub{E}$ ground state. 
In both models, $ t / U\!\! =\!\! 0.1 $, $ eR E / U\!\! =\!\! 2.5 \times 10^{-2} $, and 
$\alpha\!\! =\!\! 0.1 $.
(c) Dependence of $\Delta_{21}$ on the amplitude E, for 
${\bf E}\parallel y$ and $ eRE_{max} / U \!\! = \!\! 2.5 \times 10^{-2} $.
Filled (empty) symbols refer to the A (B) model, squares 
(triangles) correspond to $\alpha\!\!=\!\! 0.1$ ($\alpha\!\!=\!\!0$) two-site 
contributions.}
 \end{center} 
\end{figure}
In Fig.\,\ref{fig2}(b), we show the overlap between the projected ground 
state $ | \Psi_1^0 \rangle $ and the $|S=1/2, S_{12}=0,1 \rangle$
states for a given $S_z$ as function of the direction of $\vctr{E}$ (angle $\theta$).
The results coincide with the ones from
Eq.\,(\ref{eq:hespin}), for both models A and B.  
In addition, we find that the splitting ($\Delta_{21} \equiv E_2-
E_1$) between the two lowest energies varies  by less than $5\%$ with
 $\theta$, in agreement with $H\sub{E}\subup{spin}$  that predicts no $\theta$-dependence at all.  

In Fig.\,\ref{fig2}(c) we isolate the contribution to $ \Delta_{21} $ 
arising from the single- and two-site terms. All these  
contributions scale linearly with $\vctr{E}$ for every $\theta$.
The dependence of $\Delta_{21}$ on $t$, however, is model-dependent. 
In particular, in model A, the contributions to $\Delta_{21} $ arising 
from the single- and two-site terms scale like $ (t/U)^3$ and $ (t/U) $,
{\resp}  Analogous power-law dependences are found in model B, where the 
single-site (two-site) contribution scales as $(t/U)^4$ ($(t/U)^3$),
and two-site terms dominate in both models when $t\ll U$.
Additional mechanisms, such as the relative displacements of the ions, can contribute to the coupling between spin and electric field.

{\textit{Spin coupling to cavity electric fields.}}  Exchange coupling of SMMs has
been demonstrated in dimers \cite{werns02}.  The use of this short-range and (so far)
untunable interaction requires additional resources for quantum information processing \cite{troiani05a}. Efficient spin-electric  interaction, on the other hand, provides a
route to long-range and switchable coupling between SMM qubits.  In particular,
microwave  cavities are suitable for reaching the 
strong-coupling regime for various  qubit systems 
\cite{WSB+04,BI06,ADD+06,TGL08}.
Here, we propose to use such cavities to control  single SMMs and, moreover, to couple the spin-qubits of distant SMMs placed inside the same cavity.

The  interaction of a single SMM with the 
cavity field reads, $\delta H\sub{E}= d \,
 \vctr{E}_0'\cdot \vctr{C}_{\parallel} (b^{\dagger}+b)$,
where $\vctr{E}_0'$ is the
rotated electric field of amplitude
$|\vctr{E}_0|\propto\sqrt{\hbar\omega/{\mathcal{V}}}$ 
inside the cavity of volume ${\mathcal{V}}$ \cite{BI06}, and $b$ is the
annihilation operator for the photon mode of frequency $\omega$.  In the rotating wave approximation, the low-energy Hamiltonian
of $N$ SMMs interacting with the cavity mode is
$H\sub{s-ph}\equiv \sum_{j}H^{(j)}+\omega\, b^{\dagger}b$, where %
\begin{equation}
H^{(j)}=\Delta \sub{SO} \mathcal{C}_z^{(j)}S_z^{(j)}+{\bf
B}\cdot\bar{\bar{g}}\cdot{\bf
S}^{(j)}+dE_0(e^{i\varphi_j}b^{\dagger}\mathcal{C}_{-}^{(j)}+\hc),
\label{HamCav}
\end{equation}
with $\mathcal{C}_{\pm}^{(j)}=\mathcal{C}_{x}^{(j)}\pm
i\mathcal{C}_y^{(j)}$ and $\varphi_j=\theta_j + 7\pi/6$.
Here, $H\sub{s-ph}$ reduces to
the well-known Tavis-Cummings model \cite{haroche} when the spins
are in eigenstates of $S_z^{(i)}$, and  $\vctr{B}\parallel z$.  However, if  $\vctr{B}\nparallel z$ it is possible to couple both  chiralities and  total spins of  distant molecules.  
Typically, the electric fields in cavities are
weaker, $|\vctr{E}_{0}|\approx 1\unit{V/cm}$ for $\hbar\omega\approx
0.1\unit{meV}$ \cite{TGL08}, than the ones near STM tips, thus giving $\tau
\sub{Rabi} \approx 0.01-100\unit{\mu s}$. Obviously, decreasing the cavity volume ${\mathcal{V}}$
would give shorter $\tau\sub{Rabi}$.  Coupling of distant
SMMs can be controlled by tuning two given molecules in and out of
resonance with the cavity mode, e.g., by applying additional local
electric fields.  Further effects such as the state transfer between
stationary and flying qubits, or the SMM-photon entanglement, can be
observed in a system described by $H\sub{s-ph}$. 

In conclusion, we find an exchange-based mechanism that couples
electric fields to spins in triangular molecular antiferromagnets.
While our results are derived for Cu$_3$, analogous
symmetry arguments are expected to apply to other molecular magnets
that lack inversion symmetry, such as V$_{15}$\cite{chiorescuV15},
Co$_3$\cite{corronadoCobalt2},
Dy$_3$\cite{powellsessolli08}, Mn$_{12}$\cite{FST+96,TLB+96,BKW00}, {\it etc.}

We thank D. Klauser, M. Affronte and V. Bellini for useful discussions. 
We acknowledge financial
support from the Swiss NSF, the NCCR Nanoscience Basel;
the Italian MIUR under FIRB Contract No. RBIN01EY74;
the EU under "MagMaNet" and "QuEMolNa".

\end{document}